\documentclass[doublecol]{epl2}
\usepackage{soul}
\usepackage{graphicx}
\usepackage{dcolumn}
\usepackage{bm}
\usepackage{amsmath}
\usepackage{amssymb}
\usepackage{bbold}

\title{Thermodynamic arrow of time of quantum projective measurements}
\author{Juyeon Yi\inst{1}\thanks{E-mail: \email{jyi@pusan.ac.kr}} \and Beom Jun Kim\inst{2}\thanks{E-mail: \email{beomjun@skku.edu}}}
\shortauthor{J. Yi and  B. J. Kim}

\institute{
  \inst{1} Department of Physics, Pusan National University, Busan 609-735, Korea \\
  \inst{2} Department of Physics and BK21 Physics Research Division, Sungkyunkwan University, Suwon 440-746, Korea \\
}


\abstract{
We investigate a thermodynamic arrow associated with quantum projective measurements in terms of the Jensen-Shannon divergence
between the probability distribution of energy change caused by
 the measurements and its time reversal counterpart. Two physical quantities appear to govern the asymptotic values of the time asymmetry.
For an initial equilibrium ensemble prepared at a high temperature, the energy fluctuations determine the convergence of the time asymmetry approaching zero.
At low temperatures, finite survival probability of the ground state limits the time asymmetry to be less than $\ln 2$.
We illustrate our results for a concrete system and discuss the fixed point of the time asymmetry in the limit of
infinitely repeated projections. 
}

\begin{document}

\maketitle
\section{Introduction}
The arrow of time is the directionality or the asymmetry of time, which presents oddness if one runs a video clip of an irreversible physical process backward.
While all the microscopic laws of physics are entirely symmetric with respect to the direction of time, the second law of thermodynamics proclaims that the entropy of any isolated system must at least increase, and this entropy increase is responsible for the time asymmetry. The quantum dynamics governed by the Schr\"{o}dinger or the Heisenberg
equation is also time symmetric. In quantum theory, the measurement is a crucial cause for the direction of time~\cite{landau,schulman}. Standard theory of quantum mechanics postulates that upon the action of measurement, the state of a measured system collapses probabilistically onto one of the eigenvectors of the associated observable~\cite{neumann}.
In the presence of such state reductions, the system dynamics is nonunitary and hence no longer reversible in time.

Many approaches have been put forth to understand how irreversibility emerges from the time symmetric physical laws~\cite{zeh,review}. What matters is to devise a proper measure with not only a forward running clock but also a backward running clock. Recent studies on the fluctuation theorems provide insightful views in this aspect. It was suggested that the time asymmetry of a thermodynamic process can be measured by the Jensen-Shannon~(JS) divergence
between two probability distributions of the work acquired from a forward process and from
its time reversed process~\cite{feng, J}. The quantum arrow of time is commonly presumed identical to the thermodynamic arrow of time, and yet, in order to assign the thermodynamic arrow explicitly to the process of quantum measurements
it must be natural to consider its consequences on the same ground as thermodynamic processes.

In this work, we study the thermodynamic arrow of quantum projective measurement in the framework of fluctuation theorem for the nonequilibrium work~\cite{jarzynski, crooks}. We assume that a system departs from an equilibrium state with the thermodynamic arrow of zero length; in the stationary state, the time's arrow is invisible. Projective measurements not commuting with the system Hamiltonian are performed on the system and cause the energy change, $ \varepsilon = E_{f}-E_{i}$, where $E_{i}$ and $E_{f}$ denote the system energy before and after the measurements, respectively. This energy change is called the work in usual considerations where it is induced by applying a time-dependent potential~\cite{tasaki,monai}.
We consider the probability distribution of the energy change, $p(\varepsilon)$, as a key quantity reflecting the
consequence of measurements. In that, a dissimilarity between $p(\varepsilon)$
and $p(-\varepsilon)$ manifests the time asymmetry,
which can be efficiently
measured by the JS divergence. The time asymmetry of the quantum measurements in this scheme crucially depends on the initial temperature of the system to be measured. It is found that at high temperatures JS divergence vanishes with increasing temperature and quantum measurements become time symmetric at infinite temperature.
 In view of a conventional idea that quantum effects hardly survive at high temperature, the vanishing time asymmetry of measurements done on the maximally mixed ensemble at infinite temperature is obvious. However, as will be shown, the high-temperature behavior of the time asymmetry is in fact determined by the energy dispersion in an {\it infinite} temperature ensemble. On the other hand, time asymmetry becomes enhanced at low temperatures but limited less than its maximum value by the survival probability of the ground state.
We take a concrete example, and illustrate the behaviors.

\section{System}
Let us suppose a system initially in equilibrium
with a thermal reservoir at temperature $T$. The distribution of the system energy is described by the density
 matrix, $\rho^{eq}_{i}=e^{-\beta {\cal H}}/Z$ with $1/\beta =k_{B}T$ and the canonical partition function, $Z=\mbox{Tr}e^{-\beta {\cal H}}$ for the system Hamiltonian ${\cal H}$. Disconnecting the system from the reservoir and leaving
  the system Hamiltonian intact, we let the system evolve in the presence of measurements.
We consider that the measurements are performed every time $\tau$ for a certain time interval $(\tau, {\cal T})$: The first measurement is done at a time $\tau$ and followed by subsequent measurements up to the last ($M$th)measurement at a time ${\cal T}=M\tau$. In the standard scheme of quantum mechanics, the measurements is described by the projection operator, $P_{\alpha}=|\alpha\rangle \langle \alpha|$, where
$|\alpha\rangle$ is the eigenstate of an observable, $A=\sum_{\alpha}a_{\alpha}P_{\alpha}$ with $a_{\alpha}$ being
the eigenvalue of the observable.
The time evolution of the state is then determined by a combined action of the unitary time evolution and projections to give the corresponding time evolution operator,
\begin{equation}\label{timeevol}
\overline{U}_{\{\alpha\}}({\cal T})=P_{\alpha_{M}}U(\tau)\cdots P_{\alpha_{2}}U(\tau)P_{\alpha_{1}}U(\tau),
\end{equation}
where the time independent Hamiltonian generates the unitary time evolution operator, $U(\tau)=\exp(-i{\cal H}\tau/\hbar)$, for a time interval $\tau$ between consecutive projections.

The energy change of the system caused by the projective measurements
is determined by the transition probability from the initial energy eigenstate with eigenvalue $E_{i}$ to a final state
with energy value $E_{f}$:
\begin{equation}\label{tran}
\Gamma_{f,i}^{\{\alpha\}}=|\langle E_{f}|\overline{U}_{\{\alpha\}}|E_{i}\rangle|^{2},
\end{equation}
which depends on the sequence of the measurement outcome, $\{\alpha\} \equiv \{\alpha_{M},\alpha_{M-1},\cdots,\alpha_{1}\}$.
The probability distribution function~(pdf) of the energy change, $E_{f}-E_{i}\equiv \varepsilon$, is then given by
\begin{equation}\label{pdf1}
p(\varepsilon)=\sum_{E_{f},E_{i}}\sum_{\{\alpha\}}\delta (\varepsilon-E_{f}+E_{i})\Gamma^{\{\alpha\}}_{f,i}p^{eq}(E_{i}),
\end{equation}
where $\sum_{\{\alpha\}}$ denotes the summation over all the possible sequences of the measurement outcomes realized in each running of the protocol, and $p^{eq}(E_{i})=e^{-\beta E_{i}}/Z$ is the canonical distribution of the initial energy.
Equivalent description can be obtained also by using characteristic function
\begin{equation}\label{ft}
G(u)=\int_{-\infty}^{\infty}d\varepsilon e^{iu\varepsilon}p(\varepsilon) ,
\end{equation}
 which generates the $n$th cummulants of the energy change through
 ${\cal C}_{n}=(-i)^{n}[\partial \ln G(u)/\partial u ]_{u=0}$.
Inserting Eq.~(\ref{pdf1}) into Eq.~(\ref{ft}), we find that the characteristic function can be written as
\begin{equation}\label{cha}
G(u)=Z^{-1}\sum_{\{\alpha\}}\mbox{Tr} e^{iu {\cal H}}{\overline U}_{\{\alpha\}}({\cal T}) e^{-(iu+\beta) {\cal H}} {\overline U}^{\dagger}_{\{\alpha\}}({\cal T}).
\end{equation}
Note here that the characteristic function (\ref{cha}) can be written in terms of intermediate transition probabilities
$\gamma_{n,n-1}\equiv |\langle \alpha_{n}|U(\tau)|\alpha_{n-1}\rangle|^{2}$:
\begin{equation} \nonumber
ZG(u)=\sum_{\{\alpha\}} \langle \alpha_M |e^{iu{\cal H}}|\alpha_{M}\rangle \langle \alpha_{1}|e^{-(iu+\beta){\cal H}}|\alpha_{1}\rangle \prod_{n=1}^{M-1}\gamma_{n+1,n}.
\end{equation}
Substituting $u=-u'+i\beta$ and replacing the dummy variables $\alpha_{n}$ with $\alpha_{M-n+1}$, we find
\begin{equation}
ZG(-u'+i\beta)=ZG(u'),
\end{equation}
the Fourier transform of which directly leads to a symmetry relation between $p(\varepsilon)$ and $p(-\varepsilon)$ as
\begin{equation}\label{crooksm}
p(\varepsilon)=e^{\beta \varepsilon}p(-\varepsilon).
\end{equation}

Let us add a remark on the connection of this result to the fluctuation theorems. First, the Crooks
relation for the nonequilibrium work is given by $p_{f}(w)=e^{-\beta \Delta F}e^{\beta w}p_{b}(-w)$, where $p_{f}(w)$ and $p_{b}(w)$ denote the probability distribution function of the work $w$ acquired from the forward
protocol, and from the backward protocol, respectively~\cite{crooks, tasaki}. Here $\Delta F$ represents the free
energy change between the initial equilibrium with initial Hamiltonian ${\cal H}$ and the final equilibrium with ${\cal H}'$.
Since in our case the system Hamiltonian remains constant, the forward and the backward protocol to prescribe the
 paths of the Hamiltonian variance along the time axis are identical to each other, and therefore should give the same probability distributions as $p_{f}(w)=p_{b}(w)$. Furthermore, the projective measurements alone do not lead any new
 equilibrium states, and hence no free energy change occurs; $\Delta F=0$. Putting these together, we find that Eq.~(\ref{crooksm}) is the Crooks relation for the energy change induced by projective measurements. Also, integrating Eq.~(\ref{crooksm}), we find a sum rule, $\langle e^{-\beta \varepsilon}\rangle =1$ with
 $\langle \cdots \rangle$ denoting the average taken with respect to $p(\varepsilon)$. This is the Jarzynski identity with $\Delta F = 0$, endowing energy changes induced by the measurements with the meaning of a nonequilibrium work.

\section{Time asymmetry}
Let us now examine the time asymmetry associated in the energy change by the projective measurements.
Consider a mixed distribution, ${\cal P}(\varepsilon)=[p(\varepsilon)+p(\varepsilon^{R})]/2$, where $p(\varepsilon^{R})$ represents the probability to find a time reversed energy change, $\varepsilon^{R}=-\varepsilon$. In addition, a binary variable $z$ is introduced as an indicator to tell whether a sampled $\varepsilon$ is from $p(\varepsilon)$ or from its time reversal counterpart, $p(-\varepsilon)$. We assign $z=1$ if $\varepsilon$ is sampled from $p(\varepsilon)$ and $z=0$, otherwise. Consider now the joint probability of the two variables $P(\varepsilon,z)$. If the measurements are totally time symmetric to yield $p(\varepsilon)=p(-\varepsilon)$, the statistics of $z$ and $\varepsilon$ must be independent from each other. In this case, the joint probability becomes $P(\varepsilon,z)={\cal P}(\varepsilon)q(z)$ with the probability of the binary variable $q(z)=1/2$. On the other hand, if time asymmetry associated with the measurements is strong to result in $p(\varepsilon)$ very different from $p(-\varepsilon)$, such factorization of the joint probability is no longer valid. Therefore, the mutual dependence between $\varepsilon$ and $z$ can be a measure for the time asymmetry, which can be quantified by~\cite{tm}
\begin{equation}\label{mut}
I(\varepsilon;z)=\sum_{z=0}^{1}\int d\varepsilon P(\varepsilon,z)
\ln\left[\frac{P(\varepsilon,z)}{{\cal P}(\varepsilon)q(z)}\right].
\end{equation}
As it appears, $I$ vanishes if $z$ is statistically independent from $\varepsilon$. For correlated $z$ and $\varepsilon$, $I$ is finite and becomes large as the dissimilarity between $P(\varepsilon,z)$ and ${\cal P}(\varepsilon)q(z)$ increases.

Further using $P(\varepsilon,1)=p(\varepsilon)/2$ and $P(\varepsilon,0)=p(-\varepsilon)/2$, one can find that Eq.~(\ref{mut}) becomes the Jensen-Shannon distance between $p(\varepsilon)$ and $p(-\varepsilon)$~\cite{grosse}:
\begin{equation}
I(\varepsilon;z)=D[p(\varepsilon)\parallel {\cal M}]+D[p(-\varepsilon)\parallel {\cal M}]
\end{equation}
with $D(p\parallel q)=(1/2)\int dx p(x)\ln[p(x)/q(x)]$ being the Kullback-Leibler divergence.
JS divergence is a true metric and well defined even for a discrete distribution, which is not the case for all
information measures, for example, the Kullback-Leibler divergence. Not only that, its value falls in the range
from 0 to $\ln 2$, providing a physical interpretation as a length of time's arrow~\cite{feng}. When JS divergence
becomes zero for $p(\varepsilon)=p(-\varepsilon)$, time's arrow is of zero length and therefore invisible. On the other hand, if $p(\varepsilon)$ is perfectly separated
from $p(-\varepsilon)$, JS divergence is given by $\ln 2$ corresponding to a single bit of information on the time direction.

Using the property, $p(\varepsilon)=e^{\beta \varepsilon}p(-\varepsilon)$ given in Eq.~(\ref{crooksm}), we find that the JS divergence is written as
\begin{equation}\label{a}
I(\varepsilon;z)=\int_{-\infty}^{\infty} d\varepsilon p(\varepsilon)\ln \left[\frac{2}{1+e^{-\beta \varepsilon}}\right]\equiv A,
\end{equation}
which we refer to as time asymmetry $A$.
The inverse temperature $\beta$ plays a key role in determining the time asymmetry, as it appears not only
 in the logarithm but also in the distribution function $p(\varepsilon)$ given in Eq.~(\ref{pdf1}).

As mentioned earlier, one can see that at an infinite temperature~($\beta =0$) the time asymmetry vanishes. The asymptotic behavior at a  high temperature can be written in terms of series expansion of $\beta$, where the coefficients are given by the derivatives of $A$ with respect to $\beta$ evaluated at $\beta =0$. Taking derivative of Eq.~(\ref{a}) with respect to $\beta$, we obtain
 \begin{equation}\label{tempdep1}
\frac{\partial A}{\partial \beta}=-\int_{-\infty}^{\infty}d\varepsilon \frac{\partial p(\varepsilon)}{\partial \beta}\ln (1+e^{-\beta \varepsilon})
+\int_{-\infty}^{\infty}d\varepsilon \frac{\varepsilon p(\varepsilon)}{1+e^{\beta \varepsilon}}.
\end{equation}
The second integral vanishes because the integrand $ \varepsilon p(\varepsilon)/(1+e^{\beta \varepsilon})$ is an odd function of $\varepsilon$ due to the Crooks relation Eq.~(\ref{crooksm}). At $\beta =0$, the first term in Eq.~(\ref{tempdep1}) also vanishes for the normalized $p(\varepsilon)$:
\begin{equation}
\left(\frac{\partial A}{\partial \beta}\right)_{\beta =0} \propto \frac{\partial}{\partial \beta}\int_{-\infty}^{\infty} d\varepsilon p(\varepsilon) = 0.
\end{equation}
On the other hand, the second derivative of the time asymmetry evaluated at $\beta=0$ is a nonvanishing quantity to give $A\approx C\beta^{2}$ with
\begin{eqnarray}\label{aasymp}
C&=&\frac{1}{2}\left(\frac{\partial^{2} A}{\partial \beta^{2}}\right)_{\beta=0}=
\frac{1}{4}\left[\frac{\partial }{\partial\beta}\int d\varepsilon \varepsilon p(\varepsilon)\right]_{\beta = 0} \\ \nonumber
&=& \frac{1}{4}\left(\frac{\partial \langle \varepsilon\rangle}{\partial \beta}\right)_{\beta =0}.
\end{eqnarray}

The average of the energy change, or the first cummulant, $\langle \varepsilon\rangle =(-i)[\partial \ln G(u)/\partial u]_{u=0}$, is given by an average energy difference:
\begin{equation}\label{workave}
\langle \varepsilon\rangle = \langle {\cal H}_{H}\rangle_{eq}-\langle {\cal H}\rangle_{eq},
\end{equation}
where ${\cal H}_{H}$ denotes the Hamiltonian at time ${\cal T}$ in the Heisenberg picture for the time evolution operator, Eq.~(\ref{timeevol}), ${\cal H}_{H}= \sum_{\{\alpha\}}{\overline U}^{\dagger}_{\{\alpha\}}({\cal T}){\cal H} {\overline U}_{\{\alpha\}}({\cal T})$. The average of a quantity $X$ with respect to the initial equilibrium distribution is defined as $\mbox{Tr}X\rho_{i}^{eq}\equiv \langle X\rangle_{eq}$ with $\rho_{i}^{eq}=e^{-\beta {\cal H}}/Z$.
At infinite temperature, since $\rho_{i}^{eq}=\mathbb{1}/D$ with $D$ being the Hilbert space dimension, an equality, $\langle {\cal H}_{H}\rangle_{eq}=\langle {\cal H}\rangle_{eq}$, holds and consequently, the average of the energy change vanishes: $\langle \varepsilon\rangle =0$.
However, the rate of change of $\langle \varepsilon \rangle$ with respect to $\beta$ is a nonzero quantity even at
infinite temperature. Note that the derivative of the second term in Eq.~(\ref{workave}), the internal energy at the initial equilibrium state, amounts to the energy fluctuation: $-\partial \langle {\cal H}\rangle_{eq}/\partial \beta = \langle {\cal H}^{2}\rangle_{eq}-\langle {\cal H}\rangle_{eq}^{2}$. The first term in Eq.~(\ref{workave}) on the other hand is the average energy of the system in the nonequilibrium state at the end of the measurement protocol, and its derivative
with respect to $\beta$ is given by $-\partial \langle {\cal H}_{H}\rangle_{eq}/\partial \beta = \langle {\cal H}_{H}{\cal H}\rangle_{eq}-\langle {\cal H}\rangle_{eq}\langle {\cal H}_{H}\rangle_{eq}$, which is a correlation function between ${\cal H}_{H}$ and ${\cal H}$. Collecting theses, we obtain
\begin{equation}\label{corr}
\left(\frac{\partial \langle \varepsilon \rangle}{\partial \beta}\right )_{\beta =0}=\langle {\cal H}^{2}\rangle_{eq,\beta=0},
-\langle {\cal H}_{H}{\cal H}\rangle_{eq,\beta=0}.
\end{equation}
This is in fact directly related to the dispersion of $\varepsilon$. Note that the second moment of the energy difference is given by
\begin{equation}\label{endis}
\langle \varepsilon^{2}\rangle = \langle {\cal H}^{(2)}_{H}\rangle_{eq}-2\langle {\cal H}_{H}{\cal H}\rangle_{eq}+\langle {\cal H}^{2}\rangle_{eq}
\end{equation}
with the Heisenberg operator of the squared Hamiltonian, ${\cal H}^{(2)}_{H} =  \sum_{\{\alpha\}}{\overline U}^{\dagger}_{\{\alpha\}}({\cal T}) {\cal H}^{2} {\overline U}_{\{\alpha\}}({\cal T})$. For $\langle \varepsilon\rangle =0$ and $\langle {\cal H}^{(2)}_{H}\rangle_{eq} = \langle {\cal H}^{2}\rangle_{eq}$ at $\beta=0$, we find that the dispersion of $\varepsilon$ at infinite temperature is identical to Eq.~(\ref{corr}) apart from
a proportional constant:
\begin{equation}\label{rel}
\sigma^{2}_{\beta=0}=\langle \varepsilon^{2}\rangle_{\beta=0}=
2 \left(\frac{\partial \langle \varepsilon\rangle}{\partial \beta}\right)_{\beta = 0}.
\end{equation}

It might be thought that in the infinite-temperature limit, quantum projective measurements gives null influence on the system dynamics, and yields $p(\varepsilon)=\delta (\varepsilon)$. This is however not the case, as can be seen in Eq.~(\ref{pdf1}), and more easily
 from the characteristic function, Eq.~(\ref{cha}). For $\beta \rightarrow 0$, obviously $G(u)\neq 1$, indicating that $p(\varepsilon)\neq \delta(\varepsilon)$. Instead, $G(u)=G(-u)$ leads to $p(\varepsilon)=p(-\varepsilon)$. Such symmetric probability distribution yields a vanishing average, but can have finite fluctuations of $\varepsilon$. We have shown that the dispersion of $\varepsilon$ determines the asymptotic behaviors of the time asymmetry at high temperature. As given in Eq.~(\ref{endis}), the correlation between energies before and after the projective measurements emerges as an important physical quantity. It can be expected that a number of projections demolish the memory of inherent dynamics of the system, and correspondingly the correlation vanishes to yield
 $\langle {\cal H}_{H} {\cal H}\rangle_{eq} \approx \langle {\cal H}_{H}\rangle_{eq}\langle {\cal H}\rangle_{eq}$ for $M\gg 1$. In the case, the dispersion of the energy change at $\beta =0$ simply reads the equilibrium energy fluctuation of the system. Later we discuss this behavior
 in more details by taking an example system.

Let us now examine the low-temperature limit. Inserting the expression for the work pdf, Eq.~(\ref{pdf1}) into
Eq.~(\ref{a}) and expanding the logarithmic function, $\ln(1+e^{-\beta w})$, as a series of $e^{-\beta w}$, we obtain the following expression of the time asymmetry:
\begin{equation}\label{js2}
A=\ln 2 +\sum_{E_{f},E_{i}}\Gamma_{f,i}p^{eq}(E_{i})\sum_{n=1}^{\infty}\frac{(-1)^{n}}{n}e^{-n\beta(E_{f}-E_{i})}
\end{equation}
 with $\sum_{\{\alpha\}}\Gamma^{\{\alpha\}}_{f,i} \equiv \Gamma_{f,i}$.
At very low temperatures, the contributions from final energy values satisfying $E_{f} > E_{i}$ in the summation becomes strongly suppressed. Further since the system is most likely in the ground state with the energy $E_{gs}$, only taking into account contribution by $E_{f}=E_{i}=E_{gs}$, we reach the approximate form of the time asymmetry at low temperature:
\begin{equation}\label{lowtempa}
A\approx (1-\Gamma)\ln 2,
\end{equation}
where the factor $\Gamma$ is the survival probability that the system is still found
in the ground state after a series of projections:
\begin{equation}\label{sur}
\Gamma = \sum_{\{\alpha\}} |\langle E_{gs}|\overline{U}_{\{\alpha\}}({\cal T})|E_{gs}\rangle|^{2} .
\end{equation}
One can find that as long as $\Gamma$ is finite, the time asymmetry of the projective measurements is less than its maximum value, $\ln 2$. The survival probability of the initial state, $\Gamma$, depends on the dynamic details involved in the state evolution described by Eq.~(\ref{timeevol}). In particular, the role of the time interval $\tau$ between consecutive measurements is well demonstrated in the quantum Zeno effect~\cite{qzeno1, qzeno2, qzeno3} that
frequent measurements slow down the evolution of a quantum system. This effect takes an important place in the theory of quantum measurements as a direct consequence of von Neumann's picture of quantum measurements as an instantaneous projection. In that, the survival probability is a key quantity, and its experimental observations in optical setups~\cite{zenoexp1, zenoexp2} has directly verified the theoretical predictions. Our finding reveals another physical significance of the survival probability to quantify the time asymmetry of the quantum measurements through Eq.~(\ref{lowtempa}).
\begin{figure}[t]
\includegraphics[width=.4\textwidth]{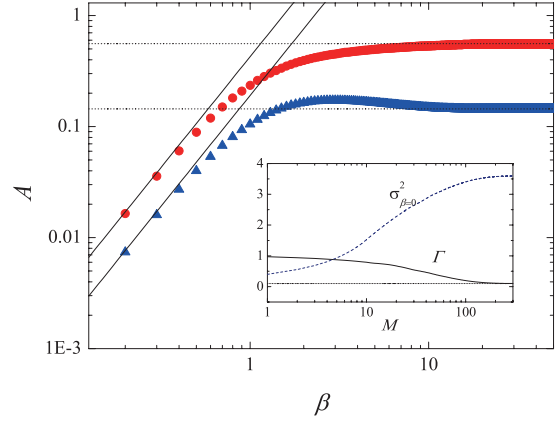}
\caption{Time asymmetry of the binary measurement, Eq.~(\ref{pro}), of a particle occupancy in a one-dimensional chain as a function of the inverse temperature, where we compare the different measurement numbers, $M=10$~(triangles) and $M=100$~(circles). The solid lines represent the high-temperature behavior, $A=\sigma^{2}_{\beta =0}\beta^{2}/8$, and
the dotted lines are given by Eq.~(\ref{lowtempa}). Here the dispersion of the energy change,
$\sigma^{2}_{\beta=0}$, and the survival probabilities, $\Gamma$, corresponding to $M=10$ and $M=100$ can be found in the inset figure displaying their $M$ dependence.}
\end{figure}

\section{Example}
Let us now illustrate the behaviors of the time asymmetry for a concrete system: A particle moving on a one-dimensional chain, which in the tight-binding
picture is described by the Hamiltonian,
\begin{equation}\label{ham}
H=-\gamma \sum_{n=1}^{N}(|n\rangle \langle n+1|  +|n+1\rangle \langle n|),
\end{equation}
where $\gamma$ denotes the hopping strength. We let $\gamma =1$ and use the hopping strength as an energy unit in presenting results. Suppose a specific measurement to observe the occupancy of a certain tight-binding site, say, $|1\rangle$, which can be represented by the projection
operators,
\begin{equation}\label{pro}
P_{1}=|1\rangle \langle 1|, \hspace{0.3cm} P_{0}=\sum_{n=2}^{N}|n\rangle \langle n|,
\end{equation}
where $P_{0}$ represents the projection onto the state where the particle occupies one of the $N-1$ sites other than the site $1$.
These projection operators are complete, $P_{1}+P_{0}=1$ and idempotent $P_{i}P_{j}=\delta_{i,j}$ for $i=0,1$.
The projection, $P_{0}$, is degenerate $\mbox{Tr}P_{0}=N-1$, while $\mbox{Tr}P_{1}=1$, giving the Hilbert
space dimension, $N = \mbox{Tr}(P_{0}+P_{1})$.

In order to examine the time asymmetry Eq.~(\ref{a}) for this model, it is essential to obtain the probability distribution function $p(\varepsilon)$ given in Eq.~(\ref{pdf1}). We consider explicit expression of the factor   $\sum_{\{\alpha\}}\Gamma^{\{\alpha\}}_{f,i}$ entering Eq.(\ref{pdf1}).
First note that in the case of a single measurement~($M=1$), we have
\begin{eqnarray}
\sum_{\{\alpha\}}\Gamma^{\{\alpha\}}_{f,i} &=&\sum_{\alpha_{1}=0}^{1}
\langle E_{f}|P_{\alpha_{1}}U|E_{i}\rangle
\langle E_{i}|U^{\dagger}P_{\alpha_{1}} | E_{f}\rangle  \\ \nonumber
&\equiv &\langle E_{f}|X^{1}|E_{f}\rangle,
\end{eqnarray}
with the operator $X^{1}$
\begin{equation}\label{m1}
X^{1} = P_{0}U|E_{i}\rangle \langle E_{i}|U^{\dagger}P_{0} +  P_{1}U|E_{i}\rangle \langle E_{i}|U^{\dagger}P_{1}.
\end{equation}
Let us here define a linear map ${\cal M}$ as
\begin{eqnarray}\label{map}
X^{m+1} &=& P_{0}UX_{m}U^{\dagger}P_{0} +  P_{1}UX_{m}U^{\dagger}P_{1} \\ \nonumber
&=& {\cal M}X^{m},
\end{eqnarray}
and then Eq.~(\ref{m1}) corresponds to $X^{1}={\cal M} X^{0}$ for the initial operator $X_{0}\equiv |E_{i}\rangle \langle E_{i}|$.
If $M=2$, the transition probability factor becomes
\begin{eqnarray}
\sum_{\{\alpha\}}\Gamma^{\{\alpha\}}_{f,i} &=&\sum_{\alpha_{2}=0}^{1}\sum_{\alpha_{1}=0}^{1}
\langle E_{f}|P_{\alpha_{2}}UP_{\alpha_{1}}U|E_{i}\rangle \\ \nonumber
&&\times \langle E_{i}|U^{\dagger}P_{\alpha_{1}}U^{\dagger}P_{\alpha_{2}} | E_{f}\rangle  \\ \nonumber
&=& \sum_{\alpha_{2}=0}^{1} \langle E_{f}|P_{\alpha_{2}}U X^{1}U^{\dagger}P_{\alpha_{2}}|E_{f}\rangle \\ \nonumber
&\equiv & \langle E_{f}| X^{2}| E_{f}\rangle,
\end{eqnarray}
where $X^{2}$ is obtained from the two-fold applications of the mapping (\ref{map}), $X^{2}={\cal M}^{2}X_{0}$.
Continuing in this way, one can notice that for general $M$  the transition amplitude in
Eqs.~(\ref{tran}) and (\ref{pdf1}) can be written as
\begin{equation}\label{tran2}
\sum_{\{\alpha\}}\Gamma_{f,i}^{\{\alpha \}} = \langle E_{f} | X^{M} |E_{f}\rangle ,
\end{equation}
with $X^{M}={\cal M}^{M}X^{0}$.

The iterative transformation Eq.~(\ref{map}) is performed in two steps.
The matrix element $Y^m_{ij} \equiv \langle i | U X^{m} U^\dagger | j \rangle$
is first computed, and then for the binary projection (\ref{pro})  an elimination procedure is done in such a way
that $X^{m+1}_{1j} = X^{m+1}_{j1}=0$ if $j\neq 1$, and  $X^{m+1}_{ij}=Y^m_{ij}$ otherwise.
As result of $M$-fold application of this transformation, we obtain the transition amplitude (\ref{tran2}) and  accordingly $p(\varepsilon)$. It is then straightforward to calculate quantities derived from the pdf such as the time asymmetry $A$ in Eq.~(\ref{a}) and the energy dispersion in Eq.~(\ref{endis}).

In Fig.~1, we present the time asymmetry as a function of the inverse temperature, $\beta$. In the high-temperature regime~($\beta \lesssim 1$),
the time asymmetry is very weak and follows the asymptotic line given by $A\approx \sigma^{2}_{\beta=0}\beta^{2}/8$ (the solid lines), as noted in Eqs.~(\ref{aasymp}) and (\ref{rel}). As temperature decreases~(increasing $\beta$), the time asymmetry becomes enhanced, and saturates into the value of Eq.~(\ref{lowtempa}) with the survival probabilities $\Gamma$ corresponding to $M=10$ and $M=100$~(the dotted lines). The $M$-dependence of $\Gamma$ is shown in the inset figure, along with,
$\sigma^{2}_{\beta=0}$, the dispersion of the energy change for the infinite-temperature initial state.
It can be seen that as the projections are repeatedly performed, the survival
probability gradually decreases and reaches the saturation $\Gamma \approx 1/10$ for $M\gtrsim 200$.
In several studies~\cite{flores,gordon,yi}, a universal fixed point of a quantum state under repeated projective measurements was reported, where the survival probability is shown to be $1/D$ for $D$ denoting the Hilbert space dimension. Therefore, in our case, the time asymmetry at low temperatures is given by $A=(1-1/N)\ln 2$ for
the survival probability of the ground state $\Gamma =1/N$, as demonstrated in Fig.~1.
Meanwhile, the dispersion of the energy change also exhibits the saturation behavior~(see the inset figure) for large $M$, which is again the consequence of the fixed point of the survival probability. Recalling that the correlation between energies after and before the measurements is given
by $\partial \langle {\cal H}_{H}\rangle_{eq}/\partial \beta$, as noted around Eq.~(\ref{corr}),
we find that the correlation function vanishes in the infinite-measurement limit for which the density matrix approaches the equal probability state as $\sum_{\{\alpha \}}{\overline U}_{\{\alpha\}}\rho^{eq}_{i}{\overline U}^{\dagger}_{\{\alpha\}}=\mathbb{1} / N$ for $M\rightarrow \infty$,
and therefore, $\langle {\cal H}_{H}\rangle_{eq}$ becomes temperature independent.  In this case, $\sigma_{\beta =0}^{2}$ given in Eq.~(\ref{rel}) together
with Eq.~(\ref{corr}) is solely determined by the energy fluctuation due to an identity
$\langle {\cal H}_{H}\rangle_{eq,\beta =0}= \langle {\cal H}\rangle_{eq,\beta=0}$. For the present example, one can obtain the dispersion
as $\sigma_{\beta=0}^{2}=2\sum_{n}E^{2}_{n}/N - 2(\sum_{n}E_{n}/N)^{2}$ with $E_{n}=-2\cos[n\pi/(N+1)]$ denoting the energy levels of
the one-dimensional chain, where the summation runs from $n=1$ to $n=N$. For $N=10$, $\sigma_{\beta=0}^{2}\approx 3.6$ for $M\gtrsim 200$, as displayed in the inset of Fig.~1. The uniqueness of the fixed point of equal probability state and the convergence rate critically
depends on the time interval between measurements and the Hamiltonian governing the system
dynamics. For the detailed account of this subject, we refer the reader to a recent work~\cite{yi}.

\section{Summary}
In summary, we proposed a measure for thermodynamic arrow associated with quantum projective measurements, adopting the framework of
fluctuation theorems. We considered the probability distribution of energy change caused by the measurement, $p(\varepsilon)$ and
found the Crooks relation exists, viz. $p(\varepsilon)=e^{\beta \varepsilon}p(-\varepsilon)$, where
the inverse temperature of the initial equilibrium state, $\beta$, appears to determine the relative weight between $p(\varepsilon)$
and its time reversal counterpart, $p(-\varepsilon)$. The time asymmetry is then measured by JS divergence between
these two probability distributions, which becomes zero if $p(\varepsilon)=p(-\varepsilon)$, and $\ln 2$ if
$p(\varepsilon) \neq p(-\varepsilon)$ everywhere. The asymptotic values of the time asymmetry are ruled by two physical quantities:
The energy fluctuations at infinite temperature and the survival probability of the ground state.
In particular, we noted that even for a maximally mixed ensemble at infinite temperature a consequence of quantum projective measurement
is revealed to cause energy fluctuations of a system. We exemplified our results in a one dimensional system, considering a binary
measurement of a particle occupancy, and discussed the fixed point of the time asymmetry.

\acknowledgments
J.Y. acknowledges support from the National Research Foundation of Korea (NRF) grant funded by the Korea government (MEST)
(Grant No.  2011-0021296), and  B.J.K. was supported by Grant No. 2011-0015731.

\end{document}